\documentclass[a4paper]{jpconf}
\usepackage{graphicx}
\begin{document}
\title{Properties of nano-graphite ribbons with zigzag edges
 -- Difference between odd and even legs --}
\author{H. Yoshioka and S. Higashibata}

\address{Department of Physics, Nara Women's University, Nara 630-8506, Japan}

\ead{h-yoshi@cc.nara-wu.ac.jp}

\begin{abstract}
Persistent currents and transport properties are investigated
 for the nano-graphite ribbons with zigzag shaped edges 
with paying attention to system length $L$ dependence. 
It is found that both
the persistent current in the isolated ring and the conductance 
of the system connected to the perfect leads show the remarkable $L$
 dependences.
In addition, the dependences for the systems with odd legs and those with even
 legs are different from each other.   
On the persistent current, 
the amplitude for the cases with odd legs shows power-low behavior as $L^{-N}$ with
 $N$ being the number of legs, 
 whereas the maximum of it decreases exponentially for the cases with even legs. 
The conductance per one spin normalized by $e^2/h$ behaves as follows.  
In the even legs cases, it decays as $L^{-2}$, 
whereas it reaches to unity for $L \to \infty$ in the odd legs
 cases. 
Thus, the material is shown to have a remarkable property that 
there is the qualitative difference between the systems with odd
 legs and those with even legs  even in the absence of the electron-electron interaction.

\end{abstract}

\section{Introduction}

Recently, graphite-based one-dimensional materials with nano-meter sizes
have been attracting much attention in both the fundamental and applied
sciences. 
A nano-graphite ribbon (NGR) is a nano-meter size graphite fragment and
is known to have a band structure strongly depending on the shape of
edges\cite{Fujita1996JPSJ,Nakada1996PRB,Wakabayashi1999PRB}.
The NGR with zigzag shaped edges, which is
schematically shown in
Figure 1 and called as a zigzag NGR in the following, 
has a metallic band structure independent of the
number of the legs $N$.
The energy dispersion is, however,  quite different from that of the
regular square lattice, especially near the Fermi energy.  
The asymptotic behavior near the Fermi energy is given as $E(k) \sim \pm
t (ka - \pi)^N$ where $t$ is the hopping energy between the nearest
neighbor atoms and $a$ is the lattice spacing, and  
then the Fermi velocity vanishes.  
Such a singular band structure is a manifestation of the fact 
that the states close to the Fermi energy 
are localized near the zigzag edges.  

In the present paper, we investigate the properties of the materials with such a
remarkable band structure through 
the persistent current in the isolated ring pierced by the magnetic flux shown
in Figure 2 and 
the conductance of the system attached with the perfect leads made from
the regular square lattices (see Figure 3).      
We take care of dependences 
on the sample length $L$ and on the number of the legs $N$. 
There exist already studies on the persistent current\cite{Nakamura2004PE} and the
transport properties\cite{Wakabayashi2000PRL,Akhmerov2008PRB} of the
zigzag NGR. 
To the best of our knowledge, however, the sample size dependences have
not been discussed for the former, and 
the system with the present setup has not been investigated for
the latter. 
Anomalous $L$ dependences 
are found in both quantities. 
More interestingly, the $L$ dependences of the even $N$ cases 
and those of the odd $N$ cases are qualitatively different from each other 
even in the absence of the mutual interaction.   
\begin{figure}[t]
\begin{center}
  \includegraphics[width=6.0cm]{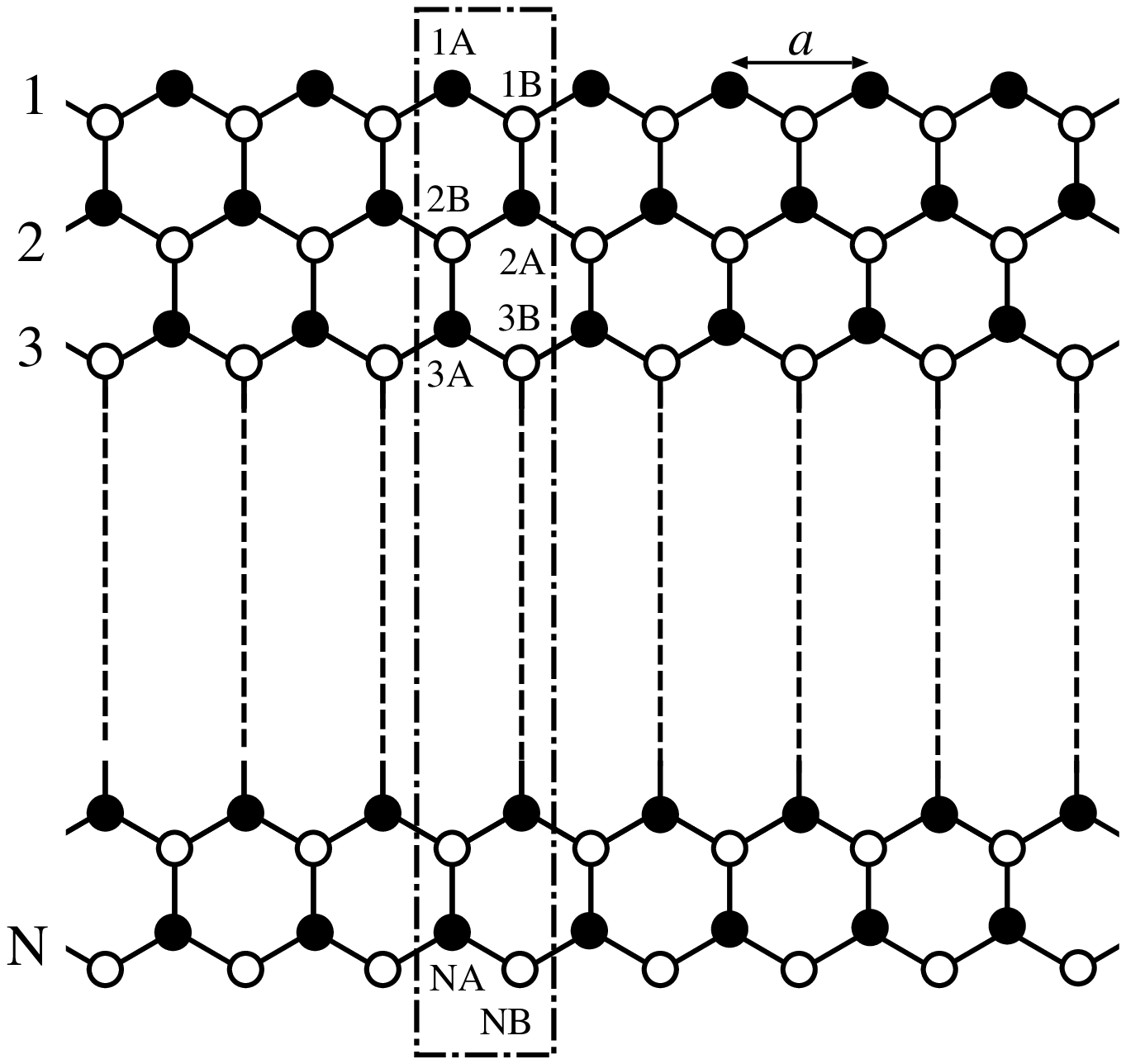}\caption{The structure of the
 zigzag NGR with $N=$ odd. 
The closed (open) circles show the A (B)
 sublattice and the rectangle with the dash-dotted line indicates a unit
 cell. Here $a$ denotes the lattice spacing.}  
\end{center}
\end{figure} 
\begin{figure}[t]
 \begin{center}
\begin{minipage}{5.7cm}
\begin{center}
  \includegraphics[width=2.5cm]{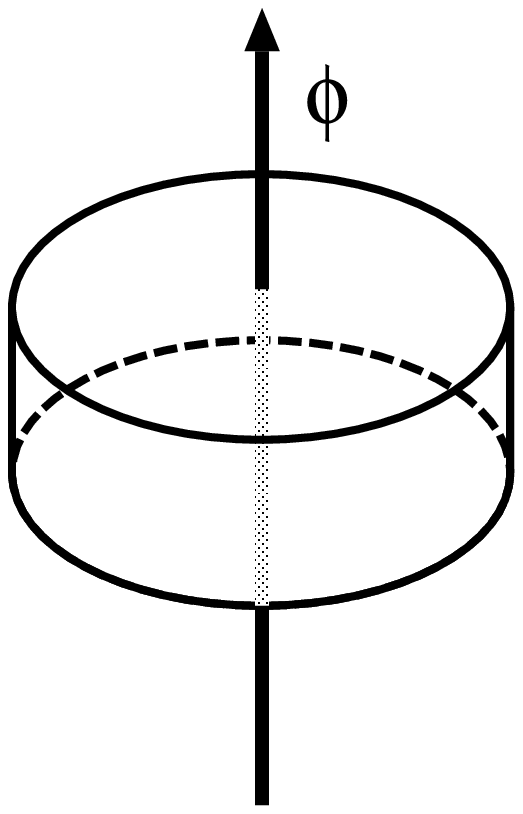}
\caption{The ring pierced by the magnetic flux $\phi$. }  
\end{center}
\end{minipage}\hspace{1pc}
\begin{minipage}{9.7cm}
\begin{center}
\vspace*{-0.7pc}
  \includegraphics[width=7.8cm]{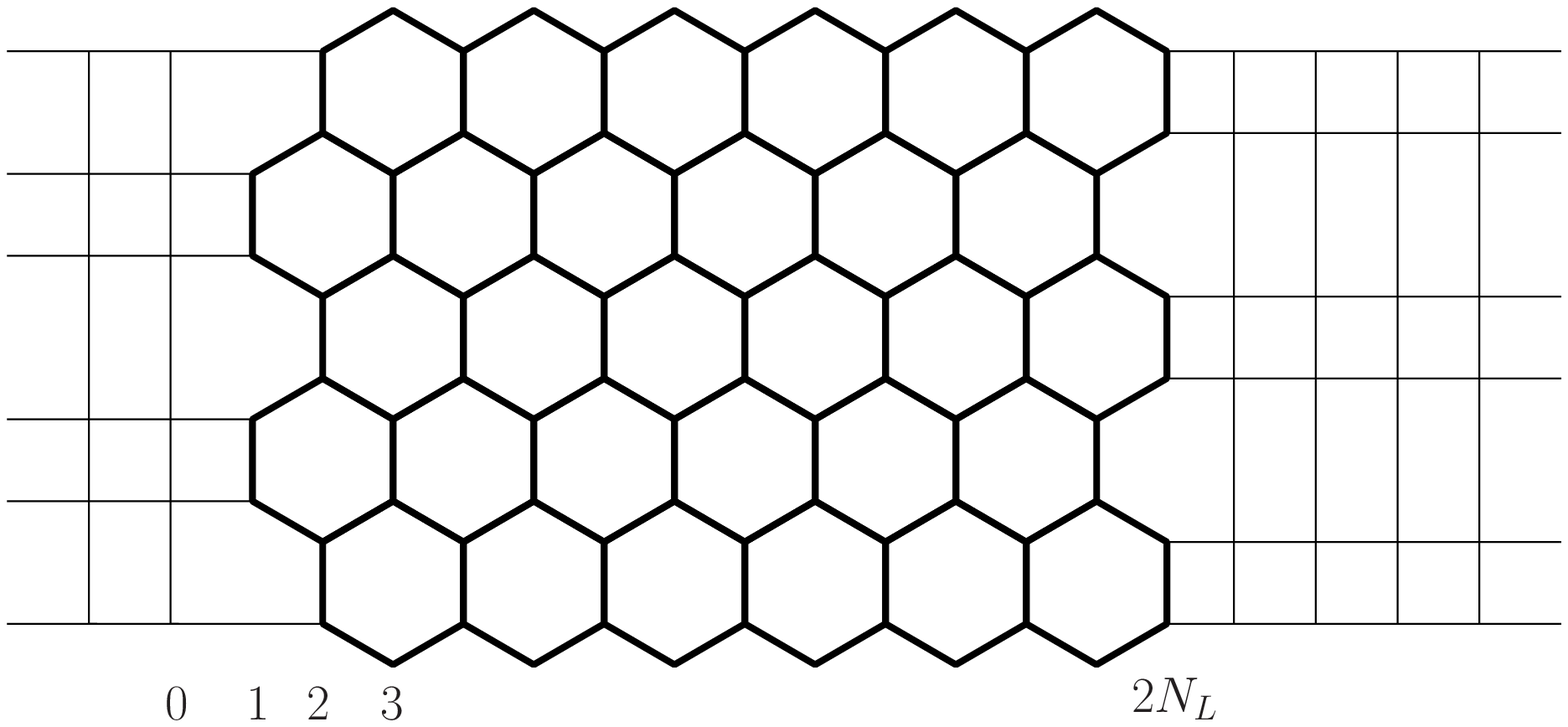}
\vspace*{-1.0pc}
\caption{The zigzag NGR attached with the perfect leads, 
which are made from the regular square lattices.}  
\end{center}
\end{minipage}
 \end{center}
\end{figure}

\section{Model and results}
We consider the zigzag NGR consisting of $N$ zigzag legs 
and the length $L = N_L a$. 
The spin degree of freedom is omitted because the
quantities we consider are independent of the spin index and taking
account of 
it makes these quantities twice larger.     
We utilize the tight-binding model with the hopping between the nearest
neighbor atoms,   
\begin{eqnarray}
 H = -t \sum_{\langle ij \rangle} \left( c_j^\dagger c_i + {\rm h.c.}
				  \right),
\label{eqn:Hamiltonian}
\end{eqnarray}
where $\langle ij \rangle$ denotes the pair of the nearest neighbor
atoms and $c_j^\dagger$ is the creation operator of the electron at the
atom $j$.    
In the following, the persistent currents and the conductances 
are discussed based on eq. (\ref{eqn:Hamiltonian})
in the absence of doping, in which case 
the Fermi level is located at the top of the valence band and at the bottom
of the conduction band.    

\subsection{Persistent currents}
The persistent current flows in the isolated rings or
cylinders made from a normal metal threaded by the magnetic flux $\phi$
without decaying.
It is a periodic function of $\phi$ with the period $\phi_0 = h/e$ 
with $h$ and $-e$ ($<0$) being the Planck constant and the electronic
charge, respectively.  
The current at the absolute zero temperature is obtained as follows, 
\begin{eqnarray}
 I = - \sum_{n,k} \frac{\partial E_n(k)}{\partial \phi},
\label{eqn:PC} 
\end{eqnarray}
where $E_n(k)$ is the energy dispersion with the subband $n$ and 
$k = 2 \pi (m+\phi/\phi_0)/L$ with $m$ being an integer.   
The summations in terms of $n$ and $k$ are performed for $E_n(k) < 0$
because of the particle-hole symmetry of the band structure.   

In the clean strict one-dimensional systems\cite{Cheung1988PRB,Cheung1988IBM}, 
the persistent current behaves like a saw as a function of the magnetic
flux and it's amplitude is given by $e v_F / L$ with $v_F$
being the Fermi velocity. 
In the case with finite width\cite{Cheung1988IBM}, the current also
shows a saw like dependences and the maximum of the amplitude 
decays as $L^{-1}$ irrespective of the number of the
legs. 
For the zigzag NGR, the persistent current can flow in spite of the
vanishing Fermi velocity\cite{Nakamura2004PE}. 
Different from the square lattices, however, 
the current shows smooth sinusoidal dependence as a function of the flux
as is shown in Figure 4 (a) and (b).   
We show the amplitude of the currents as a function of $N_L$ for $N=$
odd (c) and for $N=$ even (d) in Figure 4. 
Obviously, the odd and even legs cases are different from each other in
the $L$ dependence. 
For $N_L \gg 1$, the amplitude of the former decays as $N_L^{-N}$, 
whereas the maximum values of the amplitudes for the latter show the exponential
dependence. 
Note that we can derive the analytical expression of the amplitude for
$N=2$ as  
$I_{\rm amp}^{N=2} \propto \exp \left\{- N_L \ln[(9+\sqrt{17})/8] \right\}$
for $N_L \gg 1$, which reproduces the numerical result. 
\begin{figure}[h]
\vspace*{-1em}
\begin{center}
 \includegraphics[width=4.4cm]{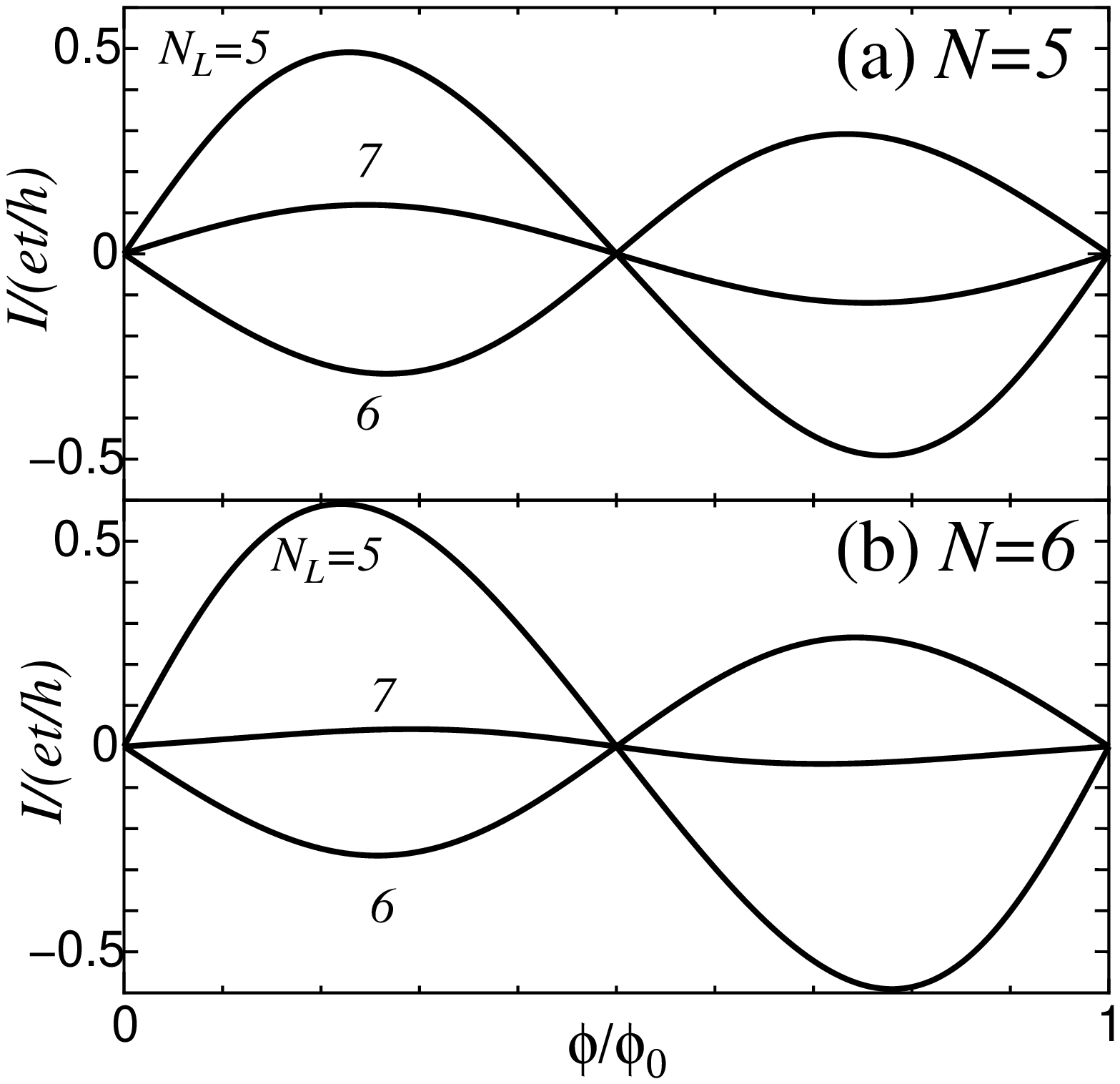}
\hspace{1.5pc}
 \includegraphics[width=4.5cm]{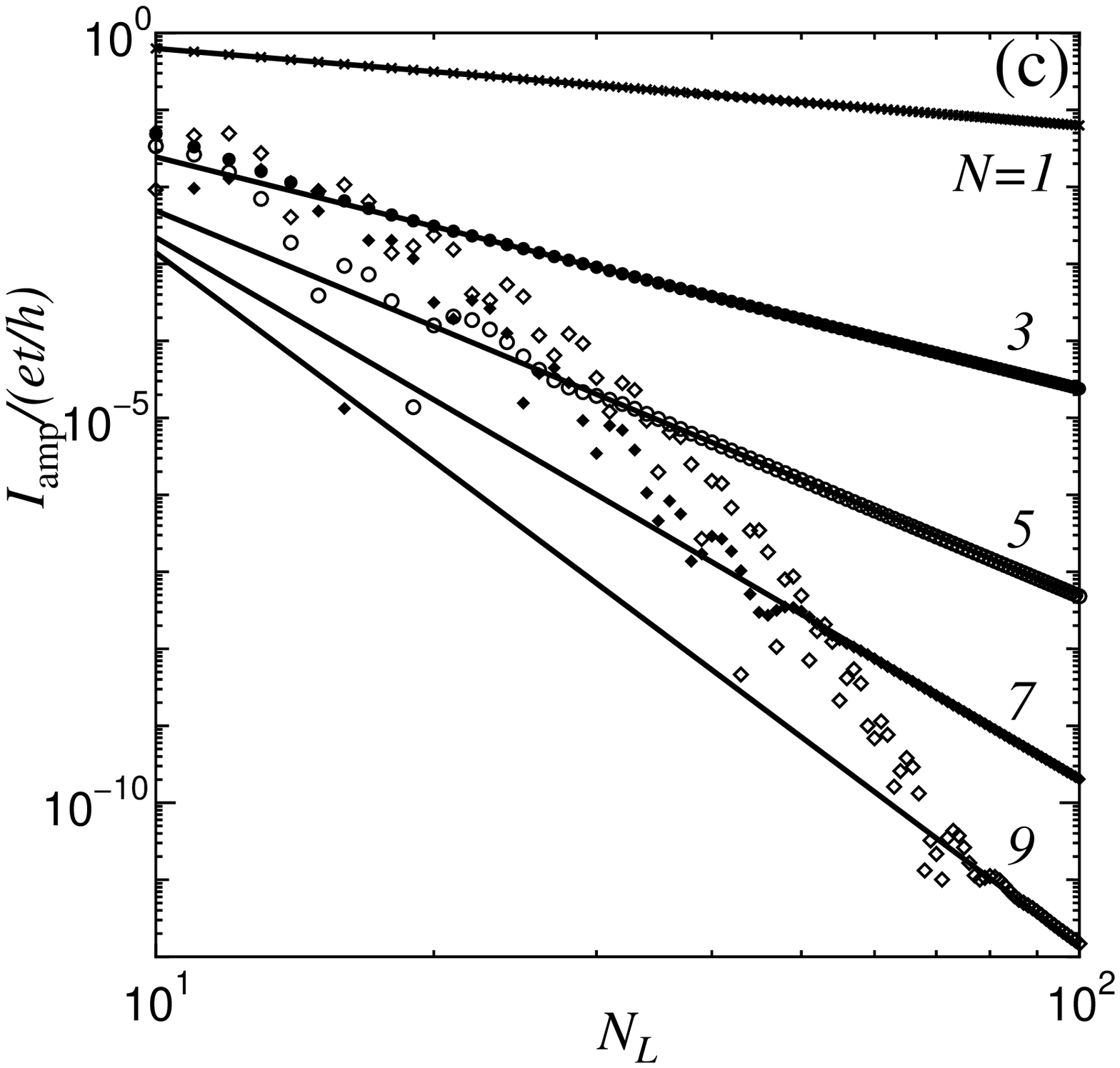}
\hspace{1.5pc}
\includegraphics[width=4.5cm]{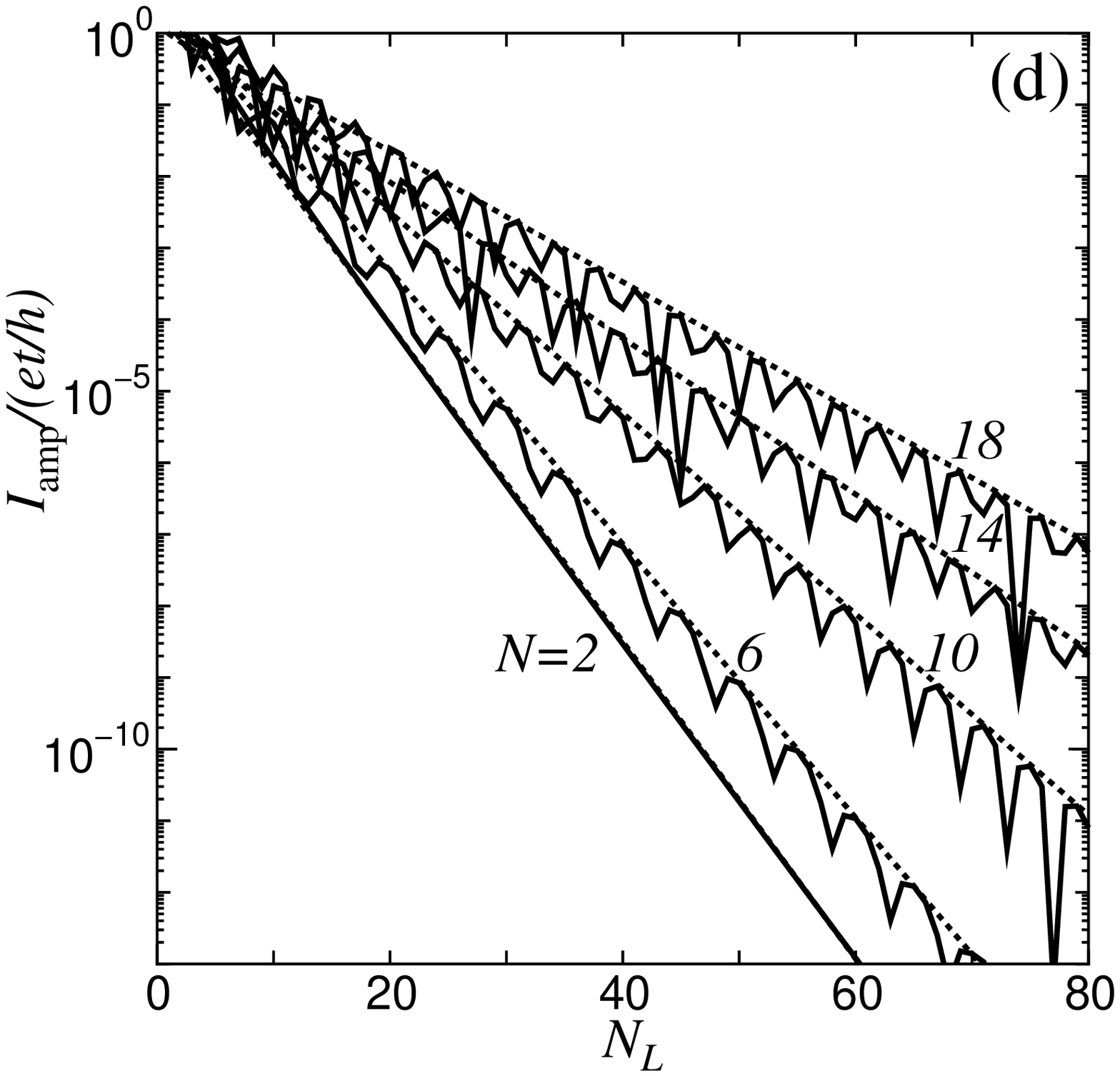}
\end{center} 
\vspace*{-1.5pc}
\caption{
The persistent currents in unit of $et/h$ as a function of the flux
for $N=5$ (a) and $N=6$ (b). 
The amplitude as a function of $N_L$ for $N=$ odd and $N=$ even are
 shown in (c) and (d), respectively.  
The solid line in each $N$ in (c) expresses the fitting by $N_L^{-N}$. 
The dotted line in each $N$ in (d) express the fitting of the maximum
 value by the exponential functions.}
\end{figure}

\subsection{Conductance}

According to Kohn\cite{Kohn1964PR}, the metallic state is defined as a state with 
the finite Drude weight $D$,  
which can be derived in the geometry in Figure 2 as 
\begin{eqnarray}
 D = - \lim_{L \to \infty} \pi L \left. \frac{\partial I}{\partial \phi}\right|_{\phi \to +0}, 
\end{eqnarray} 
where $I$ is the persistent current derived in eq. (\ref{eqn:PC}). 
In this criterion, the zigzag NGR without doping is classified into
an insulator due to the vanishing Fermi velocity in spite of  
the absence of the gap at the Fermi energy. 
Such an anomalous nature, therefore, is expected to lead 
interesting transport properties. 
We investigate the conductance of the system in Figure 3, where the
zigzag NGR is connected to the perfect leads made from square lattices, 
by utilizing the recursive green function method\cite{Ando1991PRB}. 
We note that the hopping between the nearest neighbor site in the perfect leads 
is assumed to be the same as that of the sample region for simplicity. 

\begin{figure}[h]
\vspace*{-0.4em}
\begin{center}
 \includegraphics[width=4.5cm]{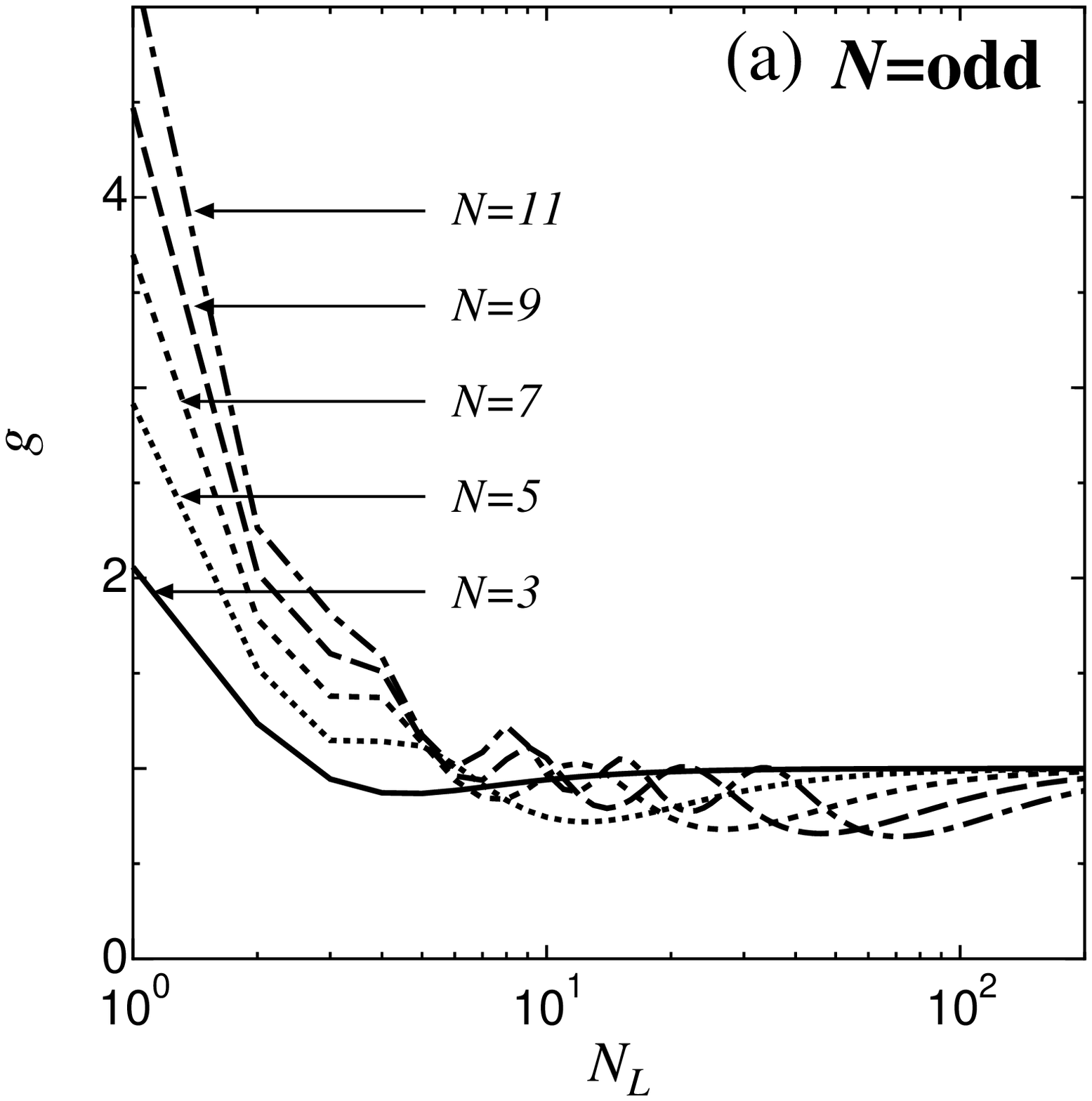}
\hspace{1.5pc}
\includegraphics[width=4.5cm]{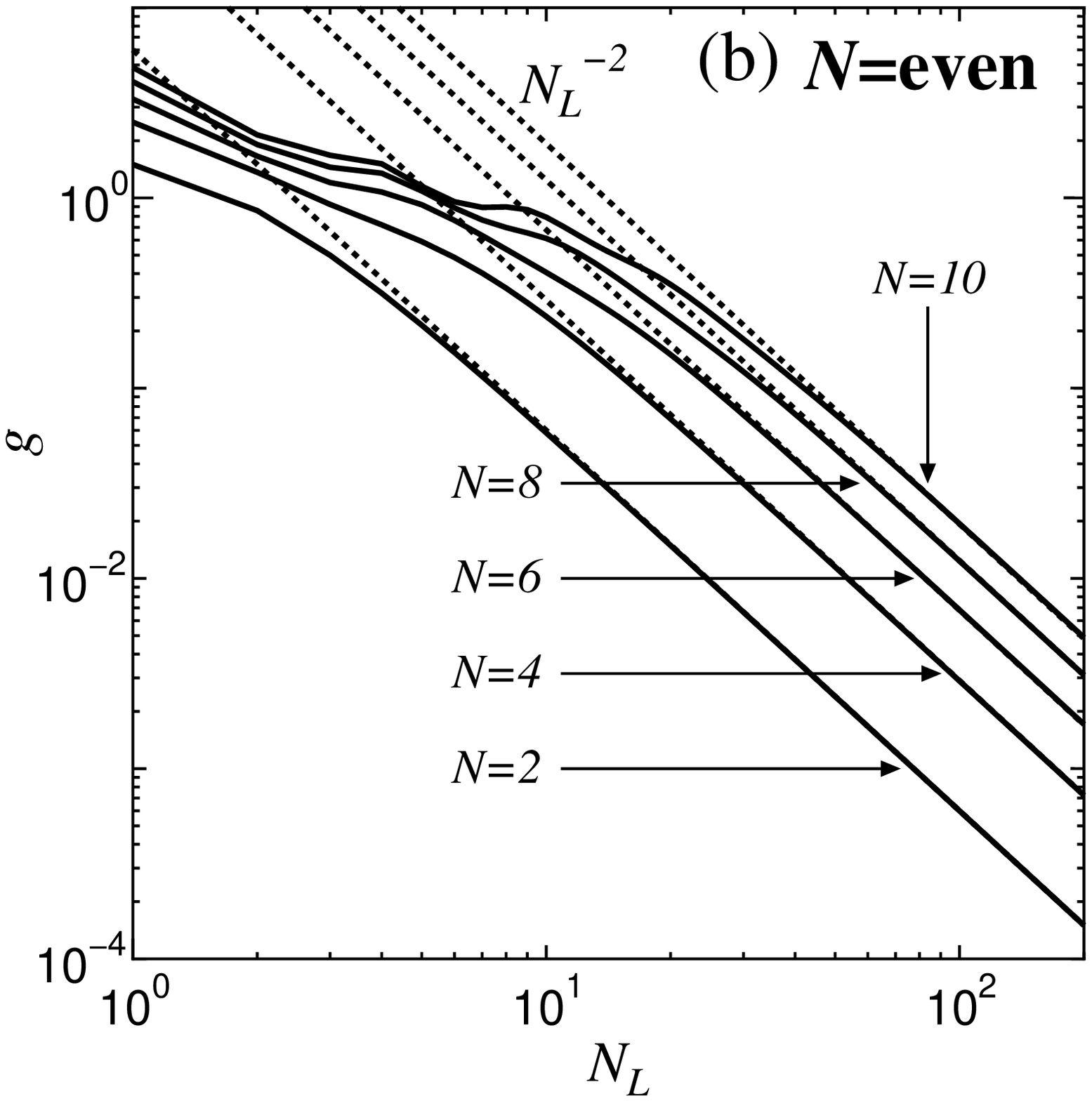}
\end{center} 
\vspace*{-1.5pc}
\caption{
The conductance of the system shown in Figure 3 as a function of $N_L$
 for $N=$ odd (a) and $N=$ even (b). }
\label{fig:g}
\end{figure}

The conductance $g$ normalized by $e^2/h$ as a function of the sample
length $N_L$ is shown in Figure \ref{fig:g} for $N=$ odd (a) and 
for $N=$ even (b).
As easily seen, the asymptotic behavior with $N_L \gg 1$ for $N=$ odd is
qualitatively different from that for $N=$ even. 
In the $N=$ odd cases, the conductance $g$ tends to unity.
On the other hand, the conductance for $N =$ even is proportional
to $N_L^{-2}$. 
Note that we can successfully obtain the analytical asymptotic ($N_L \gg 1$)
expression for $N=2$ as 
$ g^{N=2} \simeq {6}/{N_L^2}$, identical with the numerical result. 

\section{Summary and discussion} 
We theoretically investigated the properties of the
nano-graphite ribbons with zigzag shaped edges without doping through the persistent
current in the isolated ring and the conductance of the system
connected to the perfect leads. 
Both quantities are found to show the remarkable sample length $L$
dependence, which in the materials with even legs and with odd legs
are different from each other. 
It seems to be very strange that qualitative differences between the $N=$ odd
and even cases appear because we cannot find clear qualitative
discrepancy in the band structure. 
The results for $N=$ odd seem to indicate that only the band touching
the Fermi energy with the asymptotic dispersion proportional to $(ka-\pi)^N $ 
contributes to the both quantities as a whole. 
On the other hand, we suppose that  
strong suppressions of these quantities in $N=$ even compared
with $N=$ odd originate from subtle cancellation. 
In Ref. 6, Akhmerov et al. investigated the transport properties of the zigzag
NGR under the potential step and found the similar discrepancy in the
conductance between the even and odd legs cases. 
Their model is different from ours in the origin of scattering. 
In the former the origin is the potential step, whereas  
the mismatch between the square lattice and the honeycomb
lattice gives rise to the scattering in the latter. 
Further investigation is necessary for clarifying the mechanism 
leading to such strange $L$ dependences and those differences between even and
odd legs cases.


\section*{References}
\begin{thebibliography}{99}
\bibitem{Fujita1996JPSJ} Fujita M, Wakabayashi K, Nakada K and Kusakabe
	K 1996 {\it J. Phys. Soc. Jpn.} {\bf 65} 1920
\bibitem{Nakada1996PRB} Nakada K, Fujita M, Dresselhaus G and
	Gresselhause M S 1996 {\it Phys. Rev.} B {\bf 54} 17954
\bibitem{Wakabayashi1999PRB} Wakabayashi K, Fujita M, Ajiki H and
	Sigrist M 1999 {\it Phy. Rev.} B {\bf 59} 8271 
\bibitem{Nakamura2004PE} Nakamura S, Wakabayashi K, Yamashiro A and
	Harigaya K 2004 {\it Physica} E {\bf 22} 684
\bibitem{Wakabayashi2000PRL} Wakabayashi K and Sigrist M 2000 {\it
	Phys. Rev. Lett.} {\bf 84} 3390
\bibitem{Akhmerov2008PRB} Akhmerov A R, Bardarson J H, Rycerz A and
	Beenakker C W J 2008 {\it Phys. Rev.} B {\bf 77} 205416 
\bibitem{Cheung1988PRB} Cheung H F, Gefen Y, Riedel E K and Shi W H 1988
	{\it Phys. Rev.} B {\bf 37} 6050
\bibitem{Cheung1988IBM} Cheung H F, Gefen Y and Riedel E K 1988 {\it IBM
	J. RES. DEVELOP.} {\bf 32} 359
\bibitem{Kohn1964PR} Kohn W 1964 {\it Phys. Rev.} {\bf 133} A171
\bibitem{Ando1991PRB} Ando T 1991 {\it Phys. Rev.} B {\bf 44} 8017
\end{thebibliography}
\end{document}